\RequirePackage[l2tabu, orthodox]{nag}

\documentclass[12pt]{article}

\usepackage[margin=1in]{geometry}
\usepackage{latexsym}
\usepackage{amsfonts,amssymb,amsmath,amsthm,bbm}
\usepackage{graphicx, enumitem}
\usepackage[usenames,dvipsnames,svgnames,table]{xcolor}
\usepackage[all]{xy}
\usepackage{multirow}
\usepackage{float}
\usepackage{natbib}
\usepackage[left]{lineno}
\usepackage{microtype}
\usepackage{fancyvrb}
\usepackage[colorlinks=false]{hyperref}
\usepackage{bm}
\usepackage{tikz}
\usetikzlibrary{decorations.pathreplacing}

\newcounter{Problem}[section]
\newcounter{Part}[Problem]

\newcommand{\murl}[1][]{\mu_{RL#1}}

\renewcommand{\hat}{\widehat}

\newcommand{\Raise}{\raisebox{11pt}{}}

\newlength{\solutionHeight}
\newcommand{\solution}[1]{%
	\setbox1=\vbox{#1}%
	\settoheight{\solutionHeight}{\box1}%
	\vspace{\the\solutionHeight}%
	\vspace{\baselineskip}}
\renewcommand{\solution}[1]{~\par {\color{Mulberry} #1} \par \hrulefill~}

\renewcommand{\maketitle}{{\centering \Large {\makeatletter \textbf{\title} \\ \author \\ \date \par} \vspace{1.5\baselineskip}}}
\renewcommand{\title}{%
The conditionally autoregressive hidden Markov model (CarHMM): Inferring behavioural states from animal tracking data exhibiting conditional autocorrelation.
}
\renewcommand{\author}{%
Ethan Lawler$^{a,b}$, Kim Whoriskey$^{a}$, William H. Aeberhard$^{a}$, Chris Field$^{a}$, Joanna Mills Flemming$^{a}$.
}

\renewcommand{\date}{%
\today
}
\newcommand{\affiliation}{%
	$a$: Department of Mathematics and Statistics, Dalhousie University, Halifax, Canada. \\
	$b$: Corresponding author, E-mail: lawlerem@dal.ca
}

\newcommand{\comment}[1]{}

\begin{document}

\pagenumbering{gobble}

\renewcommand{\maketitle}{%
	{\makeatletter
		\noindent Title:~\title \\

		\noindent Authors:~\author \\

		\noindent \affiliation
	}
}

\maketitle

\newpage

\vspace{-2\baselineskip}
\setcounter{tocdepth}{1}
\tableofcontents

\newpage

{\centering
	\bfseries
	\title
\par}

\paragraph{Abstract}

One of the central interests of animal movement ecology is relating movement characteristics to behavioural characteristics. The traditional discrete-time statistical tool for inferring unobserved behaviours from movement data is the hidden Markov model (HMM). While the HMM is an important and powerful tool, sometimes it is not flexible enough to appropriately fit the data. Data for marine animals often exhibit conditional autocorrelation, self-dependence of the step length process which cannot be explained solely by the behavioural state, which violates one of the main assumptions of the HMM. Using a grey seal track as an example, along with multiple simulation scenarios, we motivate and develop the conditionally autoregressive hidden Markov model (CarHMM), which is a generalization of the HMM designed specifically to handle conditional autocorrelation.

In addition to introducing and examining the new CarHMM, we provide guidelines for all stages of an analysis using either an HMM or CarHMM. These include guidelines for pre-processing location data to obtain deflection angles and step lengths, model selection, and model checking. In addition to these practical guidelines, we link estimated model parameters to biologically meaningful quantities such as activity budget and residency time. We also provide interpretations of traditional ``foraging" and ``transiting" behaviours in the context of the new CarHMM parameters.

~\\


\newpage

\pagenumbering{arabic}

\section{Introduction}

The study of animal movement is fundamental to ecology because it is inherently linked to critical processes that scale from individuals to populations and communities to ecosystems \citep{Hooten::AnimalMovementTelemetryBook}. Rapid technological advancements over the past several decades have given rise to a variety of electronic tracking devices that can remotely monitor animals in challenging environments \citep{Hussey2015::aquaticAnimalTelemetry} as well as an assortment of statistical methods for analyzing the resulting (big) movement data.

Statistical models for animal movement data are most commonly formulated in discrete time \citep{Hooten2017::JABESIntro}, and are increasingly aimed at inferring behavioural ``states" from observed tracks. In this context, the data (called tracks, or location data) generally consist of a regularly observed time series of locations of an animal. Inferring behavioural states from location data was initially made possible by a proposal in \citet{Morales2004::MovementRandomWalks} to transform the data into a bivariate series of step lengths and deflection angles. In their example, they use characteristics of the step length and deflection angle series to determine when an elk is in an ``encamped" state and when it is in an ``exploratory" state.

There are many different ways to estimate behavioural states from this type of tracking data. While traditionally achieved using likelihood methods (frequentist or Bayesian), any unsupervised classification method can be used. Some examples are mixture models \citep{Morales2004::MovementRandomWalks}, clustering models, and $k$-means clustering \citep{Curry2014::movementClustering}. If researchers have actually observed an animal's behaviour at some points in time (for example through recorded video), then any supervised classification method could also be used.

The work of \citet{Morales2004::MovementRandomWalks} along with that of \citet{Jonsen::2005DCRWAnimalMovement}, popularized the state-switching model framework into the {\itshape de facto} way of analysing animal movement data in discrete time \citep{McClintock2012::MultistateRandomWalk,Whoriskey::SwitchingHMM,Patterson2017::AnimalMovementOverview}. While not all animal movement models which incorporate state-switching into the movement process have distinct behavioural states, the ones that do generally fall under the hidden Markov model (HMM) framework. These models assume that there are underlying behaviours driving the animal movement process \citep{Michelot::moveHMM}.

Hidden Markov models for animal movement have a number of desirable properties: they have an easily computable likelihood which is typically fast to optimize, the model parameters have clear interpretations, and they can fairly easily handle different types of data (including missing data) in the same model \citep{Zucchini::HMMsforTimeSeries}. The baseline formulation of the HMM has a few key assumptions: the underlying state process is assumed to form a Markov chain, and the observed step lengths and deflection angles are conditionally independent given the behavioural state.

The effect of various violations of these assumptions are discussed in \citet{Pohle::numberStatesHMM}. They found that neglecting a semi-Markov state process (which directly models state residency time), a higher order Markov chain for the behavioural process, or violations of conditional independence can introduce bias to parameter estimates and favor models which have more behavioural states than actually exist. Semi-Markov state processes are also considered in \citet{Langrock2012::HMMsExtensions} while higher order state processes are presented in \citet{Zucchini::HMMsforTimeSeries}.

The papers in the recent Journal of Agricultural, Biological, and Environmental Statistics special edition on animal movement also started to address other problems associated with the discrete-time model framework in general, such as telemetry error, irregularly spaced data, and occasional missing data \citep{McClintock::telemetryObservationError}, the temporal scale and resolution of the behaviours involved in the data \citep{Vianey::MultiScaleHMMMovement}, and choosing the number of behavioural states to use \citep{Pohle::numberStatesHMM}.
 Methods for assessing goodness of fit for animal movement models were discussed in \citet{Potts::MovementModelResiduals}, wherein they found that none of the 20 highest cited papers at the time tested goodness of fit to the data. Since then, the moveHMM \citep{Michelot::moveHMM} and momentuHMM \citep{McClintock2018::momentuHMMpackage} \texttt{R} packages have implemented easy to use residuals, although the use of residuals in the literature is still uncommon, or at least under-reported.

The current paper introduces a conditionally autoregressive hidden Markov model (CarHMM) that does not require the assumption of conditional independence of the movement process given the behavioural process. We do this by introducing an autocorrelation parameter in the step length distribution of the traditional HMM (such as that implemented in \citet{Michelot::moveHMM}). The use of an autocorrelated step length process was also present in a continuous-state model for estimating the effect of environmental covariates on behavioural memory in \citet{Forester2007::StateSpaceBehaviouralMemory}.

Throughout, we provide general practice guidelines wherever possible. Since analyses of animal movement typically use offline data , we propose standardizing the observed step lengths by dividing by the mean observed step length. This allows comparison of models across data sources, animals, species, etc. We use a lag-plot of step length for determining if the conditional autocorrelation is necessary, and note its possible use to help in choosing the number of behavioural states to use. In the case of irregularly observed tracks, we also discuss how to choose an appropriate interpolation time step for the model, as well as how to deal with extensive missing data by grouping observations.

In Section \ref{sec::modelForm}, we present the formulation of the model including computation of the likelihood and give references for the theoretical properties. Section \ref{sec::paramInterp} discusses the biology associated with specific transformations of the model parameters. Section \ref{sec::dataInspectModelCheck} deals with pre-processing the locations by choosing a time step and dealing with missing data, as well as model selection and validation.
Both Section \ref{sec::paramInterp} and Section \ref{sec::dataInspectModelCheck} are useful for most types of discrete-time models, including the HMM and the CarHMM. Section \ref{sec::Simulation} presents four short simulation studies. Section \ref{sec::mgreySeal} demonstrates best practice for using the CarHMM through the analysis of a male grey seal.

\section{CarHMM formulation}
\label{sec::modelForm}

We assume that the data consist of a set of step lengths $d_{(t,t+1)}$ between locations at time $t$ to $t+1$ and deflection angles $\theta_{t}$ between locations at times $t-1$, $t$, and $t+1$. Locations are assumed to be observed on a discrete and evenly spaced time grid. Here, step lengths measure the distance between consecutive locations, and deflection angles measure the angular change in direction between three locations. We discuss observations which are irregularly spaced in Section \ref{ssec::preProcess}.

We introduce a behavioural state process $B_{t}$ which is a Markov chain on a finite set of states $\{1,~...,~k\}$. Thus the distribution for $B_{t}$ is completely determined by the value $b_{t-1}$ of $B_{t-1}$ and the transition probability matrix $\mathbf{A}$. The $(i,j)^{\text{th}}$ entry $a_{i,j}$ of $\mathbf{A}$ gives the probability of transitioning from state $i$ to state $j$.
We assume (other choices are possible) that the initial distribution of the behavioural state Markov chain is given by the stationary distribution $\bm{\delta}$, which is the vector such that $\bm{\delta}\mathbf{A} = \bm{\delta}$ and $\sum_{i}\delta_{i}=1$.

Given the behavioural state at time $t$, the step length at time $(t,t+1)$ and deflection angle at time $t$ are assumed to be conditionally independent of all other observations and behavioural states, with the key exception that step length at time $(t,t+1)$ is allowed to be dependent on step length at time $(t-1,t)$. A first order autoregressive process is assumed for step lengths $d_{(t,t+1)}$. While any valid distributions can be used, in this presentation we assume a gamma  ($\Gamma$) distribution for step lengths and a wrapped Cauchy (WC) distribution for deflection angles $\theta_{t}$.
A $\Gamma\left[(1-\phi)\cdot\murl + \phi\cdot d_{(t-1,t)},~\sigma\right]$ distribution has mean $\mu = (1-\phi)\cdot\murl + \phi\cdot d_{(t-1,t)}>0$ (reversion level $\murl>0$, autocorrelation $0<\phi<1$) and standard deviation $\sigma>0$, with the more traditional shape and scale parameters being $(\mu / \sigma) ^ 2$ and $\sigma^{2} / \mu$, respectively.
The $\text{WC}\left(c,~\rho\right)$ distribution has center $c\in \left[-\pi,~\pi\right]$ and concentration $\rho\in\left(0,1\right)$ with density function
\[
	f\left(\theta;~c,~\rho\right) = \frac{1}{2\pi}\cdot \frac{1-\rho^{2}}{1+\rho^{2} - 2\rho\cdot\cos\left(\theta - \mu\right)}.
\]

In cases where we have all of the data before analysis (i.e.~we are not streaming the data), we standardize all step lengths by dividing by the observed mean step length. This removes units (for example, kilometers) and standardizes parameter interpretation across data sources, animals, species, etc. Comparison of standardized parameter estimates across data sources is dependent on many factors, including the temporal resolution of each data source, choices made during the model procedure, and the biology/ecology of the animals being compared. Conditional on the mean observed step length, dividing by the mean does not alter parameter inference since dividing a gamma distribution by a (non-zero) constant results in another gamma distribution. In practice, we store the observed mean step length so we can un-standardize later if desired. For the rest of the paper, we will assume that the symbol $d_{(t,t+1)}$ stands for standardized step length.

With all of the terms now defined, the CarHMM is formulated as
\[
	\renewcommand{\arraystretch}{1.7}
	\begin{array}{rll}
		\text{Location:}
			& \multicolumn{2}{l}{\mathbf{x}_{t+1} = \mathbf{x}_{t} +  d_{(t,t+1)}\cdot \mathbf{H}\left(\theta_{t}\right)\cdot\left[d^{-1}_{(t-1,t)}\cdot\left(\mathbf{x}_{t}-\mathbf{x}_{t-1}\right)\right]} \\
		\text{Action:}
			& \multicolumn{2}{l}{
				\begin{minipage}{20em}
					\[
					\renewcommand{\arraystretch}{0.8}
					\begin{array}{rl}
						d_{(t,t+1)}~\vert~B_{t}&=~b \sim \Gamma\left((1-\phi_{b})\cdot\murl[,b] + \phi_{b}\cdot d_{(t-1,t)},~\sigma_{b}\right) \\
				 		\theta_{t}~\vert~B_{t}&=~b \sim\text{WC}\left(c_{b},~\rho_{b}\right)
					\end{array}
					\]
				\end{minipage}} \\
		\text{Behaviour:}
			& \text{Pr}\left[B_{t} = j ~\vert~ B_{t-1} = i\right] = 
			  a_{ij},
			&	i,j\in\left\{1,2,...,k\right\}
			\\
		\text{Initial Conditions:}
			& \multicolumn{1}{l}{\text{Pr}\left[B_{1} = i\right] = \delta_{i}, }
			& \begin{minipage}{15em} $d_{(0,1)}$ is fixed from the data as the first observed step length. \end{minipage}\\
		%
	\end{array}
\]

Although the locations $\mathbf{x}_{t}$ themselves do not enter the likelihood for the model directly, we include the ``Movement" equation to show the connection between the locations and the step lengths and deflection angles. In this equation, $\mathbf{H}\left(\theta_{t}\right)$ represents the change in direction at time $t$. If, as we strongly recommend, the coordinates are latitude-longitude pairs, then the equation as given is more of a symbolic representation. The fully written equation is based on spherical geometry. If the coordinates are projected, then $\mathbf{H}$ can be written as a standard $2\times 2$ rotation matrix.

The likelihood is computed as the matrix product
\[
	L = \bm{\delta}\mathbf{L}\left(d_{(1,2)},\theta_{1}\right)\mathbf{A}
			\mathbf{L}\left(d_{(2,3)},\theta_{2}\right)\mathbf{A} \cdots
			\mathbf{L}\left(d_{(n,n+1)},\theta_{n}\right)\mathbf{1}_{k\times 1}
\]
where $\mathbf{L}\left(d_{(t,t+1)},\theta_{t}\right)$ is the diagonal matrix
\[
	\text{Diag}\left[\Raise f\left(\Raise d_{(t,t+1)},~\theta_{t}~\vert~B_{t} = 1\right),~...,~f\left(\Raise d_{(t,t+1)},~\theta_{t}~\vert~B_{t} = k\right) \right],
\]
and $\mathbf{1}_{k\times 1}$ is a vector of ones. Since $d_{(t,t+1)}$ and $\theta_{t}$ are considered conditionally independent given the behavioural state, their joint probability density function is the product of the individual densities. In practice, we compute the log-likelihood using forward recursion with scaling as presented in Section 3.2 of \citet{Zucchini::HMMsforTimeSeries}.

When the autocorrelation $\phi_{b}$ is fixed at 0 for all $b$, the CarHMM reduces to a standard HMM. In addition, when $\phi_{b}$ is fixed at 1 for all $b$, it is possible to show that the CarHMM reduces to a component-wise relative of the hidden Markov movement model of \citet{Whoriskey::SwitchingHMM} though the details will not be shown here. Further, other generalizations to the standard movement HMM such as adding a semi-Markov state process could be applied to the CarHMM.

When using the $\Gamma$ distribution $\murl[,b]$ must be non-negative and $\phi_{b}$ must be within the unit interval. Another useful choice of distribution for step length is the log-normal distribution, where the log of the step length has mean $(1-\phi_{b})\cdot\murl[,b] + \phi_{b} \cdot d_{(t-1,t)}$. In this case the parameters are unrestricted. However to ensure the step length process is stable, it is sufficient (but not necessary) that the estimates satisfy $\left\vert\hat\phi_{b}\right\vert <1$ for all $b$. If this is not the case it may be a sign of numerical instability in the optimizer.

We use the maximum likelihood framework; the parameters to be estimated are $\murl[,b]$, $\phi_{b}$, $\sigma_{b}$, $c_{b}$, $\rho_{b}$ for $b\in 1,~...,~k$ giving $5k$ parameters for the ``Action" distribution, and the off-diagonal transition probabilities $a_{i,j}$ for $i,j\in 1,~...,~k,~i\neq j$ giving $k\cdot(k-1)$ parameters for the ``Behaviour" distribution, for a total of $k^{2} + 4k$ parameters. The remaining transition probabilities are not free parameters since the row sums of $\mathbf{A}$ must equal 1. The unobserved behavioural states $B_{t}$ are predicted using the well known Viterbi algorithm, see e.g.~\citet{Zucchini::HMMsforTimeSeries}.

Identifiability of models in the Markov-switching autoregressive class, which includes the CarHMM, is proven in \citet{Douc::AutoregressiveMarkovRegimeNormality}, with consistency and asymptotic normality of the ML estimates when using the log-normal distribution resulting from the same paper. Consistency and asymptotic normality when using the $\Gamma$ distribution is studied in \citet{Ailliot::gammaMSAR}. One notable consistency condition is that the entries of $\mathbf{A}$ must be strictly positive for parameter estimation to be consistent. If any estimated value is close to zero, this could be a sign of having too many states in the model, unless there is a biologically meaningful reason for including the extra state.

\section{Interpretation of CarHMM parameters}
\label{sec::paramInterp}

Here we discuss the concepts of activity budget, behavioural residency time, and mean reversion level. These are all obtained as transformations of the model parameters and are related to the biology of the animal.

Both activity budget and behavioural residency time can be obtained from the transition probability matrix $\mathbf{A}$. The stationary distribution $\bm{\delta}$ itself can be interpreted as an activity budget, where the $i^{\text{th}}$ entry of $\bm{\delta}$ gives the expected proportion of time that the animal spends in the $i^{\text{th}}$ behavioural state. For example, the activity budget can give estimates of the proportion of time spent transiting as compared to foraging.

Behavioural residency time is the amount of time that an animal will remain in a given behavioural state before switching to a different state. These can be modelled explicitly using semi-Markov state processes, though in our case they follow a geometric distribution \citep{Langrock2012::HMMsExtensions}. For a geometric distribution, the expected number of time steps spent in state $i$ is given by $\bm{E}_{t}(\text{state}~i)= 1 ~/~ (1 - p_{i,i})$. When converted to real time units (hours, minutes, etc.) by multiplying by the chosen time step, this value gives an estimate of the time scale of the behaviour being modelled and is important in giving biologically meaningful interpretations to the behavioural states.

Since the step length process within a given behavioural state follows a first-order autoregressive process, the parameter $\murl[,b]$ gives the reversion level of the process for a given state. The mean reversion level is defined to be the expected value of step length as the time spent within the behavioural state approaches infinity.
\[
	\mu_{RL,b} = \lim_{t\to\infty}\bm{E}\left( D_{(t,t+1)} ~\biggr\vert B_{t'} = b ~\forall ~t' \leq t \right).
\]
Within each behavioural state this value acts as an attractor in the step length distribution. Thus consecutive step lengths in the same behavioural state will tend to converge on this value, with the strength of attraction being inversely proportional to $\phi_{b}$ and proportional to the distance of the previous step length to $\mu_{RL,b}$.

One of the attractive features of hidden Markov models for movement data is the connection between the underlying behavioural state of the Markov chain with behaviours exhibited by the animal. While the connection between the behavioural state in the model and the behaviours exhibited by the animal is sometimes tenuous, it can be useful to label the behavioural states of the model. Two common ``behaviours" used in the HMM context are ``foraging" and ``transiting" (quotations are used to emphasise that these labels may not reflect actual behaviour).

In the standard HMM, foraging is typically characterized by short step lengths and diffuse deflection angles. Transiting is typically characterized by long step lengths and deflection angles concentrated at zero degrees. We can update these behaviours in the CarHMM framework. Here foraging is characterized by short step lengths with little autocorrelation along with diffuse deflection angles. Transiting is characterized by longer and highly autocorrelated step lengths along with deflection angles concentrated at zero degrees.

\section{Data inspection and model checking}
\label{sec::dataInspectModelCheck}

\subsection{Pre-processing locations}
\label{ssec::preProcess}

When dealing with marine animal tagging data the observed locations are typically not on a regular time grid. In order to use the CarHMM, we linearly interpolate the observed locations to a regular time grid. An alternative is to use the multiple imputation approach proposed in \citet{McClintock::telemetryObservationError}. However even when using this multiple imputation approach one must choose a sensible time grid.

Interpolation requires making a few decisions such as: how are observations which are very far apart in time (long stretches of missing data) dealt with? and, what is the best time step (the time between points on the temporal grid) to use for the interpolation? We deal with the first by splitting the track into separate groups whenever the time between consecutive observations is greater than some cutoff level, which we call the group cutoff level. After defining the time step, which we require to be the same for all groups, and the group cutoff level the observed locations are interpolated within their separate groups on to the regular time grid. The interpolated locations are then processed to obtain the deflection angles and step lengths which enter the likelihood.

There are two metrics we propose to help in choosing a time step and group cutoff level: the proportional sample size
\[
	n_{\text{prop}} = \frac{\text{\# interpolated locations}}{\text{\# observed locations}}
\]
and the adjusted proportional sample size
\[
	n_{\text{adj}} = \frac{\text{\# interpolated locations} - 2 \cdot \text{\# groups}}{\text{\# observed locations} - 2}.
\]

The first is designed to preserve the number of locations in the track, while the second is designed to preserve the number of data points which enter the likelihood.

To choose a (heuristically) best time step and group cutoff level, we recommend:
\begin{enumerate}
	\item{restrict the time step to be somewhere between the median and 3$^{rd}$ quartile of the observed time differences in the original data;}
	\item{for whichever time step is chosen, restrict the group cutoff level to be no more than twice the time step (and no less than the time step itself);}
	\item{with the above restrictions, set up a grid of time steps and group cutoff levels and compute $n_{\text{prop}}$ and $n_{\text{adj}}$ for each point in the grid. Choose whichever makes both $n_{\text{prop}}$ and $n_{\text{adj}}$ as close to 1 as possible.}
\end{enumerate}

We follow these guidelines in choosing the time step for the best practice analysis of Section \ref{sec::mgreySeal}. These guidelines are meant to avoid both over-smoothing the data (and therefore losing information) by choosing too large a time step, or inadvertently replicating the data by choosing too small a time step. Values of $n_{\text{prop}}$ or $n_{\text{adj}}$ less than one are indicators of over-smoothing while values greater than one are indicators of data replication.

A brief experiment suggested that interpolation (either linear interpolation or with a variety of different splines) may introduce a significant amount of autocorrelation in the step lengths. Further investigation is needed to determine the exact effects of interpolation. Whether most of the autocorrelation is inherent to the track or is introduced by interpolating the locations, accounting for it in the model (such as the CarHMM does) is necessary.

Once the locations are interpolated and grouped with a regular time step, they are processed to obtain deflection angles and step lengths. These should be obtained from unprojected coordinates so that both the deflection angles and step lengths are accurate throughout the spatial extent of the data. The model is fitted to all of the grouped data in a single likelihood, assuming that groups are independent of each other and that the groups share the same true parameters. Within a group, the first step length of that group is taken to be the initial condition for the step length autoregressive process, and the initial distribution of the underlying state is always taken to be the stationary distribution.

\subsection{Model selection}
\label{ssec::modelSelection}

We consider two components to model selection for the CarHMM: deciding whether to use the HMM or the CarHMM (i.e., whether to fix $\phi_{b}=0$ or not), and choosing the number of behavioural states.

First we introduce the lag-plot, an exploratory graphic useful for understanding the nature of any autocorrelation present in the step length process. The lag-plot at lag $k$ is a kernel-density plot of $d_{(t,t+1)}$ against $d_{(t-k,t+1-k)}$. Examples of these plots are given in Figure \ref{fig::simCarHMMlag} of Section \ref{sec::Simulation}. These lag-plots give a more detailed description of the autocorrelation than the simple autocorrelation function, and have a couple of immediately helpful uses.

Most importanly in the current case, by examining a lag-plot at lag 1, it can be possible to determine which of the HMM and CarHMM is more appropriate for the data. These plots show the different types of autocorrelation present within the HMM and CarHMM, compared with a real dataset. The HMM plot will have a pattern of distinct circular droplets along the line $y=x$, a result of the autocorrelation in the behavioural states, while the CarHMM plot will have an elongated smear along the line $y=x$, due to the within-state autocorrelation in step lengths.

In ideal cases, the lag-plot at lag 1 may also suggest the number of states exhibited in the data. Particularly for data with HMM-like autocorrelation, the number of distinct droplets corresponds to the number of distinct states. This becomes more complicated with more latent states and with more CarHMM-like autocorrelation. Choosing the number of states to use is a notoriously difficult problem, with traditional metrics such as AIC and BIC generally selecting too many states to be biologically meaningful. For in-depth discussion, we refer the reader to \citet{Pohle::numberStatesHMM}. A recommended starting point is to use as few states as necessary to achieve an adequate fit.

Other uses of the lag-plot include comparing the autocorrelation characteristics of different time step choices. For example, we could test the intuitively attractive idea that a short time step results in step lengths with high within-state autocorrelation, while a long time step results in step lengths with low within-state autocorrelation. Further, with multi-state models the traditional autocorrelation function can do a poor job of quantifying the autocorrelation. The lag-plot includes more detail such that the characteristics of each state can be discerned.

\subsection{Residuals}
\label{ssec::residuals}

For model checking we follow the probability scale residual framework of \citet{Shepherd::residuals}, and in particular use one-step-ahead forecast residuals. Here, the forecast distribution for step length would be a mixture of gamma distributions with means $(1-\phi_{b})
\cdot\murl[,b] + \phi_{b} \cdot d_{(t-1,t)}$, standard deviations $\sigma_{b}$, and mixture rates given by the $b_{t-1}^{\text{th}}$ row of $\mathbf{A}$.

If we have specified the model structure correctly then these residuals should be uniformly distributed on (-1,1) and exhibit no autocorrelation. Further, they should have this property within each behavioural state, though small sample size can be a problem here. Departures from a uniform distribution can be detected by looking at a quantile-quantile (Q-Q) plot of the residuals. Residual autocorrelation can be identified in plots of the autocorrelation function of these residuals, or in lagplots such as those proposed for model selection.

\section{Simulation Study}
\label{sec::Simulation}

To address the performance and properties of the new CarHMM model, we present brief vignettes of four simulation studies. First, we investigate the effect of track length. Second, we look at the effect of within-state autocorrelation. The third study looks at the effect of the transition probabilities. The fourth and final study compares the regular HMM and the CarHMM. When simulating data we simulate both the underlying behavioural state and observed data from scratch. We do not reuse the behavioural states estimated from original data.

The main metric we use to assess these simulations is the interquartile range (first and third quartiles) of the state estimate error. For a particular track, the state estimate error is the percentage of state estimates which do not agree with the true simulated state. The quartiles are then computed over e.g.~50 simulations under the same model.

To account for numerical instability in the maximization of the likelihood, our fitting procedure for these studies attempts to fit the model to each simulated track at most 10 times, with different random starting values each time, until the model converges. If the model does not converge successfully on any of those 10 attempts, we remove that particular track from consideration. We also remove tracks which give unreasonable parameter estimates (in particular, any model fit which gives a stationary distribution with an entry less than 0.01), or estimates which give clear signs of numeric instability (deflection angle concentration less than $10^{-3}$, or a transition probability matrix with any row having all equal entries). When using real data we can tweak exactly how we optimize the likelihood (change various control parameters, pick starting values, etc.) so this practice is not an inherent shortcoming of the model or fitting method.

The Viterbi algorithm is currently the most common way to estimate behavioural states in HMM-like models. Briefly, the algorithm takes as input parameter point estimates and the observed data, and outputs the most likely sequence of behavioural states as a point estimate. With the Viterbi algorithm, the accuracy of the state estimates is dependent on the amount of overlap of the state-dependent distribution. If two states have significant overlap, the Viterbi algorithm will perform much worse than if the two states were distinct.

At no point are standard errors of parameter estimates or uncertainty statements about the behavioural states considered. Because of this, any error in the parameter estimates directly translates to a source of error in the state estimates, with no hope of correcting for the uncertainty in the parameter estimates. Because the Viterbi algorithm only gives the most likely sequence of behavioural states with no uncertainty estimates, the estimated behavioural states must be interpreted with care. In addition to the tenuous connection between the behavioural state labels and the biology, there is the additional problem that we do not know how likely the most likely behavioural state path is. Simulations, such as the ones below, can help determine how much error to expect. However, since the actual states are unobserved, it is not possible to know the actual error rate.

\subsection{Effect of track length}
\label{ssec::simTrackLength}

\begin{figure}
	\centering
	\includegraphics[width = \textwidth]{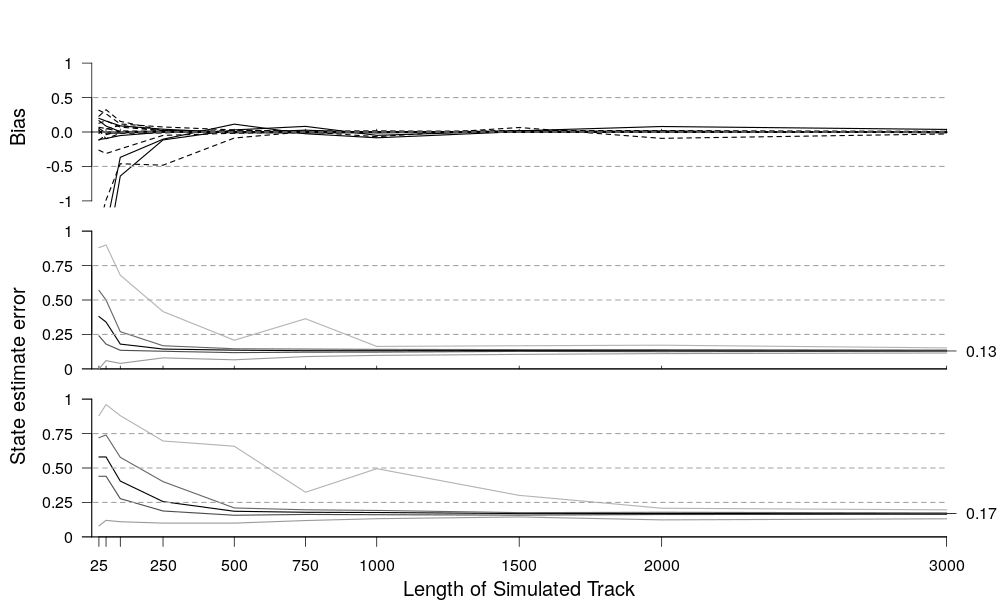}
	\caption{The top panel gives the bias for simulated tracks of different lengths under the same parameters, for both a two- and three-state model. The important feature is that the bias for all parameters converges to zero ($\sim$500 locations), showing that the parameters can be successfully estimated given a long enough track. The bottom two panels give the state estimate error give number summary (min, median, max, and quartiles). Each track length used 50 simulations. In both the two state model and the three state model, the median error rate quickly stabilizes ($\sim$250 observations for the two state model, $\sim$500 for the three state model), but does not converge to zero.}
	\label{fig::simLengthCompare}
\end{figure}

In practice, one would hope that collecting more data (i.e.~longer animal tracks) would decrease the amount of error in both parameter estimates and state estimates. Simulations suggest that, while error in parameter estimates will decrease with longer track lengths, there is an inherent amount of error to be expected in behavioural state estimates that cannot be overcome with increased track length.

Figure \ref{fig::simLengthCompare} shows the bias and state estimate error for two different models. Here we compute bias as the median difference, across simulations, of parameter estimates from the true parameter. The two state model is a HMM which takes (slightly modified) parameters estimated from an elk dataset analyzed in the vignette for the R package moveHMM \citep{Michelot::moveHMM}. The three state model is a CarHMM which takes parameters estimated from a grey seal dataset (different from the one presented in Section \ref{sec::mgreySeal}). The parameters for both models can be found in  Table \ref{tab::twoElkPars} and  Table \ref{tab::threeSealLengthPars}, respectively, of ESM \ref{sapp::simGraphs}

Figure \ref{fig::simLengthCompare} shows that collecting more and more data for a single track is not effective past a certain point. For the remaining simulation studies we use track lengths of 1,000. We report the first and third quartiles of the state estimate error, which Figure \ref{fig::simLengthCompare} suggests will have stabilized at this track length.

\subsection{Effect of autocorrelation}
\label{ssec::simPhi}

\begin{table}
	\centering
	\begin{tabular}{crccc}
		 		& 		& \multicolumn{3}{c}{State 2} \\
		 		& 		& Low									& Med 									& High 					\\
		 		& Low	& (0.131, 0.148) [4]	& (0.111, 0.128) [2]	& (0.074, 0.087) [0]	\\
		State 1	& Med	&																& (0.139, 0.161) [5]	& (0.082, 0.098) [0]	\\
				& High	&																&																& (0.209, 0.240) [0]
	\end{tabular}
	\caption{First and third quartiles for the state estimate error for different combinations of low, medium, and high autocorrelation. The number in square brackets gives the number of simulations which did not converge, out of 100 simulations.}
	\label{tab::simAutocorrelation}
\end{table}

The accuracy of the Viterbi algorithm depends heavily on the amount of overlap of the state-dependent distributions. Recall that the mean of the step length distribution is given by
\(
	(1-\phi)\cdot\murl + \phi\cdot d_{(t-1,t)}.
\)
Consider the autocorrelation $\phi$ as a weight between $\murl$, which will depend on the state at time $t$, and $d_{(t-1,t)}$, which does not. If $\phi$ is close to one in two states with drastically different $\murl$, then the two states will overlap since $\murl$ is essentially irrelevant in both states.

Table \ref{tab::simAutocorrelation} shows the state error rate for a two state model with different amounts of autocorrelation (each state taking either a low, medium, or high amount). The parameters are modified from the same elk example used in Section \ref{ssec::simTrackLength}. Overall we see that increasing the autocorrelation of both states leads to an increase in the amount of state estimate error, while differentiating the amount of autocorrelation between the two states leads to a decrease in the amount of error.

 \subsection{Effect of transition probabilities}
 \label{ssec::simTransition}

 \begin{table}
 	\centering
 	\begin{tabular}{crcccc}
 		 		& 		& \multicolumn{4}{c}{State 2~~($\phi = 0.892$)} \\
 		 		& 		& 0.5					& 0.6 					& 0.7					& 0.9 															\\
 		 		& 0.5	& (0.214, 0.234) [1]	& (0.204, 0.227) [0]	& (0.188, 0.207) [1]	& (0.125, 0.154) [2]	\\
 		State 1	& 0.6	& (0.226, 0.240) [3]	& (0.204, 0.223) [2]	& (0.194, 0.210) [0]	& (0.127, 0.147) [2]	\\
 ($\phi = 0.407$) & 0.7	& (0.224, 0.245) [2]	& (0.207, 0.223) [2]	& (0.196, 0.218) [2]	& (0.115, 0.134) [0]	\\
 				& 0.9	& (0.213, 0.242) [6]	& (0.179, 0.226) [6]	& (0.154, 0.180) [1]	& (0.082, 0.104) [2]	\\
 	\end{tabular}
 	\caption{The row and column headings give the probability of staying within the given state from one time to the next. First and third quartiles for the state estimate error. The number in square brackets gives the number of simulations which did not converge, out of 20 simulations. The amount of error decreases as the probability of remaining in state 2 increases. The error is not significantly affected by the probability of remaining in state 1.}
 	\label{tab::simTransition}
 \end{table}

Unlike in the standard HMM, the observed state-dependent distributions for the CarHMM are indirectly affected by the transition probabilities of the underlying behavioural states. States with low autocorrelation act as anchors in the step length series, while states with high autocorrelation tend to wander. The longer that an animal is in a state with high autocorrelation (by having a high probability of remaining in the same state), the more we expect that step length series to wander.

Figure \ref{fig::simTransitlag} in ESM \ref{sapp::simGraphs} gives observed step length distributions for a variety of different transition probabilities. Table \ref{tab::simTransition} gives state estimate error under the same variety of transition probabilities. We see that the probability of remaining in state 2 (with high autocorrelation) affects the state estimate error, as this probability is what determines how free the second state is to wander. The more that the high autocorrelation state drifts away from the mean of the low autocorrelation state (from left to right in the table), the less overlap there is in their distribution, which increases the accuracy of the Viterbi algorithm. The parameters can be found in  Table \ref{tab::twoSealPars} of ESM \ref{sapp::simGraphs}.

\subsection{Comparison of HMM and CarHMM}
\label{ssec::simCompare}

\begin{table}
	\centering
	\begin{tabular}{crcc}
		& & \multicolumn{2}{c}{Two State Model} \\
			\multicolumn{2}{c}{Simulated Model}	& HMM																		& CarHMM 								\\
			Fitted	& HMM		& (0.120, 0.138) [8]	& (0.434, 0.474) [0]	\\
			Model	& CarHMM	& (0.125, 0.145) [7]	& (0.072, 0.083) [0]
	\end{tabular}%
	\begin{tabular}{cc}
		\multicolumn{2}{c}{Three State Model} \\
		 	HMM																		&	 CarHMM 								\\
		 	(0.044, 0.058) [15]	& (0.373, 0.445) [30]	\\
			(0.047, 0.060) [3]	& (0.157, 0.186) [2]
	\end{tabular}
	\caption{First and third quartiles for the state estimate error. The number in square brackets gives the number of simulations which did not converge, out of 100 simulations. When the data is simulated with no within-state autocorrelation, the HMM and the CarHMM have essentially the same error rate. However, when the data is simulated with within-state autocorrelation, the HMM performs very poorly compared to the CarHMM.}
	\label{tab::simHmmvsCar}
\end{table}

\begin{figure}
	\centering
	\includegraphics[width = 0.9\textwidth]{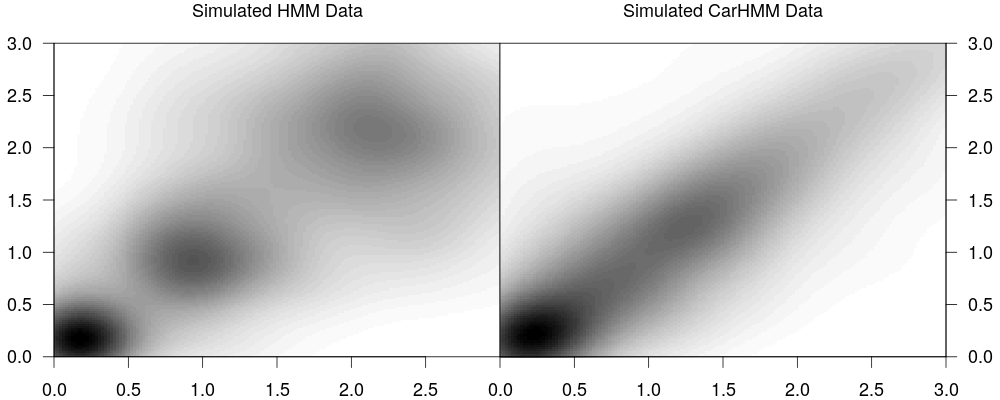}
	\caption{Lagplots for simulated HMM and CarHMM data. The three states of the HMM data are clearly shown by the three droplet patterns caused by the lack of within-state autocorrelation. The CarHMM does not clearly show the number of states, but shows the characteristic smeared line of the within-state autocorrelation. One can compare these plots to lag-plots of real data to help determine an appropriate model for the data.}
	\label{fig::simCarHMMlag}
\end{figure}

To show the importance of accounting for conditional autocorrelation in the data, we simulate data under both the HMM and the CarHMM and fit both the HMM and CarHMM to each simulation. We use parameters from two different datasets: the ``Low-High" two state parameters from the elk track considered in subsection \ref{ssec::simPhi}, and three state parameters estimated from the grey seal track analyzed in Section \ref{sec::mgreySeal}.
The parameters for the models can be found in Table \ref{tab::twoElkPars} of ESM \ref{sapp::simGraphs}, and Table \ref{tab::msealCarHMM3Pars} of Section \ref{sec::mgreySeal}, respectively. Figure \ref{fig::simCarHMMlag} shows example lag-plots under the HMM and the CarHMM for the three state model. As mentioned earlier, these plots can help in model selection.

Table \ref{tab::simHmmvsCar} shows the state estimation error rate for the four different scenarios. The CarHMM is just as effective as the HMM when fitted to HMM data with no conditional autocorrelation. However, the two-state HMM ($\sim 40-45\%$ error) performs only slightly better than random guessing ($50\%$ error) when fitted to CarHMM data with conditional autocorrelation. We expect this amount of error to persist across models that have at least one state with significant autocorrelation. The three-state HMM has the same problem, although performs much better than the $\sim$66\% error expected from random guessing.

These simulations raise interesting questions about the validity of previous research utilizing hidden Markov models with irregularly timed data, especially since we suspect a non-trivial amount of autocorrelation is introduced through interpolating the locations to a regular grid. However, we only mention this point and leave the discussion for another time.

\paragraph{Computation Time and Implementation}

All simulations were computed on a laptop running Linux with a quadcore Intel Core i7-7500U CPU with 8GB of RAM. To compare the computation speed of the HMM with the CarHMM, we timed how long it took to fit a three state HMM and a three state CarHMM 100 times to the seal data in Section \ref{sec::mgreySeal}. We also timed how long it took to simulate and refit 100 simulations from each model. The HMM averaged 2.43 seconds per fit, and an additional 0.77 seconds per simulation. The CarHMM averaged 2.37 seconds per fit, and an additional 1.23 seconds per simulation. The difference in computation time between the two models is essentially negligible. Our implementation of the CarHMM uses the R package Template Model Builder \citep{Kristensen::2015TMB}, which allows for fast computation through automatic differentiation. It also has the ability to fix parameters at given values, allowing our HMM and CarHMM implementation to be identical. Our implementation and other functional tools discussed earlier are available as an R package at the first author's GitHub page. This package also includes the data used in Section \ref{sec::mgreySeal}.

\section{Best practice analysis of a male grey seal track}
\label{sec::mgreySeal}

\begin{figure}[h!]
	\centering
	\includegraphics[width = \textwidth]{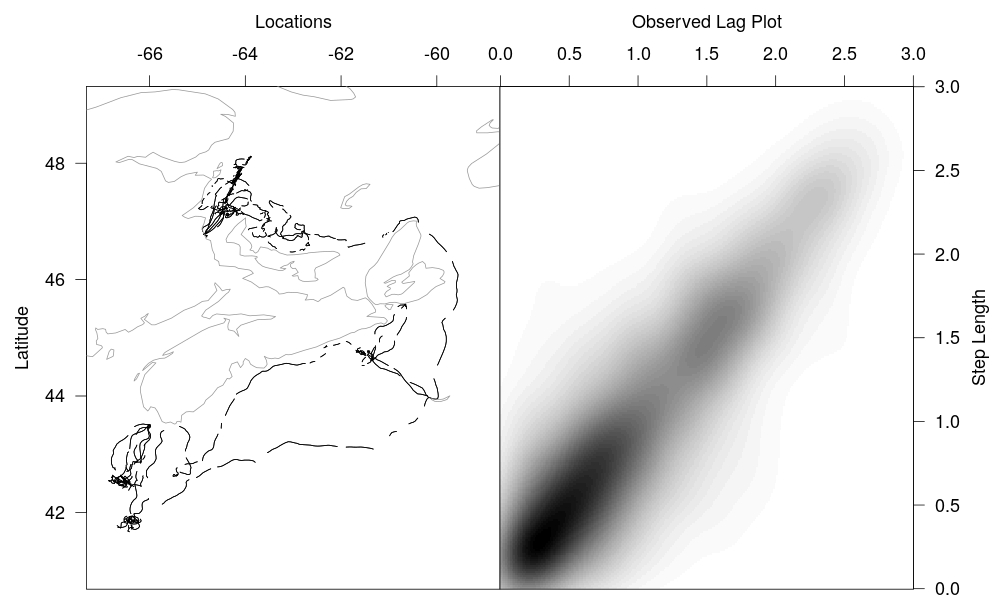}
	\caption{Map and lag plot of the grey seal track used in the best practice case study. Grey seals are large marine predators found in the North Atlantic ocean that are commonly observed travelling hundreds of kilometres to forage. This grey seal came from the Sable Island colony of Eastern Canada.}
	\label{fig::bestPracticeFigure}
\end{figure}

In this Section we demonstrate what we now consider to be basic best practice for analyzing animal movement data and reporting subsequent results. Plots, including residual plots and state estimate maps, are given in ESM \ref{sapp::msealGraphics}.

The data are a subset of a male grey seal track on the Scotian shelf, analyzed previously in \citet{Whoriskey::SwitchingHMM}. The seal was tracked using GPS with negligible observation error. Due to some data collection issues (the median time differences abruptly change without explanation) we will look at only the final 3,158 locations with time differences having a mean, median, and 3rd quartile of 100, 64, and 122 minutes, respectively.

First, one must choose values for the time step and group cutoff. To do this, set up a grid of values where the time step ranges from 60 minutes to 120 minutes by increments of 3 minutes, and the group cutoff ranges from the time step to twice the time step in increments representing 5\% of the time step. This range of values for the time step is chosen to range approximately from the median to the 3rd quartile. Refer to Section \ref{ssec::preProcess} for more detail.

Both metrics for choosing a good time step and group cutoff discussed in Section \ref{ssec::preProcess} chose an optimal time step of 66 minutes and group cutoff of 132 minutes. The resulting interpolated track consists of 3,129 locations in 251 groups with $n_{\text{prop}} = 0.991$ and $n_{\text{adj}} = 0.832$. The mean of the unstandardized step lengths is 2.10 kilometres per time step (1.91 km/hr).

The most useful plot of the data is the lag-plot of $d_{(t,t+1)}$ vs.~$d_{(t-1,t)}$ and is shown as part of Figure \ref{fig::bestPracticeFigure}. This plot shows the smeared texture that is characteristic of the CarHMM. The residuals for a two state CarHMM have autocorrelation on the border of significance. Neither a three state or a four state CarHMM give improved residuals (not shown). The data nor any of the residuals showed evidence of long-term seasonality. Given no other reason to choose a specific number of states, we recommend using the least number of states which you feel accurately describe the data.

We also remind the reader that behavioural state labels such as ``foraging" and ``transiting" may not be reflective of the actual biology.


\paragraph{Two state model} The parameter estimates are given in  Table \ref{tab::twoSealPars} in ESM \ref{sapp::simGraphs}. State 2 is interpretable as a ``transiting" behaviour. The autocorrelation parameter ($\phi_{2} = 0.89$) and concentration parameter $(\rho_{2} = 0.86)$ are suitably high, and the standard deviation $(\sigma_{2} = 0.244;~\sigma_{2}/\mu_{RL,2} = 0.14)$ is suitably low. A map of the state estimates also indicates a ``transiting" behaviour.

State 1 does not have as clear an interpretation. It may be tempting to label it a ``foraging" behaviour to complement the ``transiting" behaviour of State 2, however the parameter estimates for State 1 do not fully support this view. The autocorrelation parameter ($\phi_{1}= 0.41)$ is not close to 0 and the concentration parameter ($\rho_{1}=0.51$) is higher than expected. Further, a map of the state estimates shows that some of the behaviour picked up by this state does not have traditional ``foraging" characteristics. This suggests that State 1 may be picking up two distinct behaviours. We believe these behaviours may be a ``foraging" behaviour and a ``large area search" behaviour, although many other possibilities may exist. For this reason, we suggest using a third state to further differentiate these behaviours.

\begin{table}
	\centering
	\begin{tabular}{rcccrcccrccc}
		\multicolumn{12}{c}{Three State CarHMM Parameter Estimates} \\
		$\mathbf{d_{(t,t+1)}}$	& State 1	& State 2	& State 3 &	$\bm{\theta_{t}}$	& State 1	& State 2	& State 3	& $\mathbf{A}$				& $\mathbf{p_{\cdot,1}}$& $\mathbf{p_{\cdot,2}}$	& $\mathbf{p_{\cdot,3}}$	\\
		\hline
		$\mathbf{\bm{\mu}_{RL,b}}$	& 0.398	& 1.291		& 2.074	  & $\mathbf{c}$		& -0.129	& -0.050	& 0.002		& $\mathbf{p_{1,\cdot}}$	& 0.713					& 0.287						& 0.000					\\
		$\mathbf{\bm{\phi}_{b}}$	& 0.277	& 0.781		& 0.961	  &	$\bm{\rho}$			& ~0.402	& ~0.780	& 0.906		& $\mathbf{p_{2,\cdot}}$	& 0.149					& 0.797						& 0.054					\\
		$\bm{\sigma}$			& 0.279		& 0.318		& 0.164	  &						& 			&			&			& $\mathbf{p_{3,\cdot}}$	& 0.000					& 0.120						& 0.880					\\
								& 			& 			& 		  &					  	&			&			&			& $\bm{\delta}$				& 0.264					& 0.508						& 0.228					\\
	\end{tabular}
	\caption{Parameter estimates for a male grey seal track using the three state CarHMM.}
	\label{tab::msealCarHMM3Pars}
\end{table}

\paragraph{Three state model} The parameter estimates are given in Table \ref{tab::msealCarHMM3Pars}. State 1 is closer to a ``foraging" behaviour than it was in the two state model, and a map of the state estimates places State 1 where we might \emph{a priori} expect ``foraging" to take place based solely on the locations. State 3 is archetypal ``transiting" behaviour with both $\phi_{3}$ and $\rho_{3}$ close to 1. Based on the transition probabilities which do not allow transitions between State 1 and State 3, one would label State 2 a ``transitional" behaviour. Based on parameter estimates and a map of the state estimates there is no reason to believe that a fourth state is needed.

The expected residency times are: 3.48 timesteps (3 hr 50 min) for the ``foraging" behaviour;  4.93 timesteps (5 hr 25 min) for the ``transitional" behaviour; and 8.33 timesteps (9 hr 10 min) for the ``transiting" behaviour. The expected activity budget gives 26.4\% of the seal's time spent ``foraging", 22.8\% of its time spent ``transiting", and 50.8\% of its time transitioning between the two\comment{; the exhibited activity budget gives 30.5\% , 25.1\%,  and 44.4\%, respectively}. A simulation study of 93 convergent simulations out of 100 gave first and third quartiles of the state estimate error as $(20.5\%,~22.9\%)$.

\section{Discussion}
\label{sec::conclusion}

We have introduced the conditionally autoregressive hidden Markov model (CarHMM) for highly accurate tracking data as an alternative to both the HMM originally developed in \citet{Morales2004::MovementRandomWalks} and the HMMM documented in \citet{Whoriskey::SwitchingHMM}.

Subjective choices are often involved during data processing and model fitting. When fitting discrete-time movement models, the choice of time step often depends on the discrete behaviour of interest as well as the observation frequency \citep{Breed2012::stateSpaceTrack}. We propose a statistic to help the user choose a time step based on producing a roughly similar number of interpolated locations and data points as the original tracking data set. This could be combined with the multi-scale model of \citet{Vianey::MultiScaleHMMMovement} to study the discrete behaviour of interest. We have additionally proposed a method to deal with long periods of missing data. In some formulations of the HMM, a missing location enters the joint likelihood by including the contribution of the underlying behavioural state Markov chain ($\mathbf{A}$ in our formulation above) while removing the observation contribution for that location ($\mathbf{L}$ in our formulation) \citep{Zucchini::HMMsforTimeSeries}. We instead decided to split the track into multiple groups for compartmentalized model fitting, and offered metrics for choosing how to perform this partition. Frequent long periods of missing data can be common in marine environments.

The CarHMM draws a new link between HMMs and the DCRWS model of \citet{Jonsen::2005DCRWAnimalMovement}. Within the marine context, the two most commonly sought-after behavioural states are foraging and transiting. These states are typically assumed to follow an area-restricted search pattern, whereby foraging patches are characterized by shorter step lengths occurring in diffuse directions, and are interspersed with periods of directed travel consisting of longer step lengths directed straight ahead (see e.g.~\citet{Whoriskey::SwitchingHMM}). While these states can be directly inferred from the state-dependent distributions of the HMM, the interpretation of these state estimates resulting from the DCRWS is less straightforward. Within the DCRWS, the main parameter influencing the step lengths is an autocorrelation term ($\gamma$).
Usually (again see e.g.~\citet{Whoriskey::SwitchingHMM}), high $\gamma$  values are interpreted as highly persistent movement (indicative of transiting) and low $\gamma$ values constitute highly random movement (representing foraging). As a result, transiting and foraging are not necessarily delineated by longer and shorter step lengths. The CarHMM combines the two approaches such that we now have a clear interpretation of the step lengths but can still account for the fact that some animals will tend to move in a similar (or dissimilar) manner across time. These properties make the CarHMM a useful model for linking movement data to behavioural characteristics. \\

~\\

{\itshape Acknowledgements}
\vspace{4\baselineskip}

\bibliographystyle{apalike}
\bibliography{bibliography}

\newpage
\pagenumbering{gobble}

\appendix

\section{Extra Supplementary Material}
\label{app::graphics}

\subsection{Simulation graphics and tables}
\label{sapp::simGraphs}

\begin{figure}[H]
   \centering
   \includegraphics[width = \textwidth]{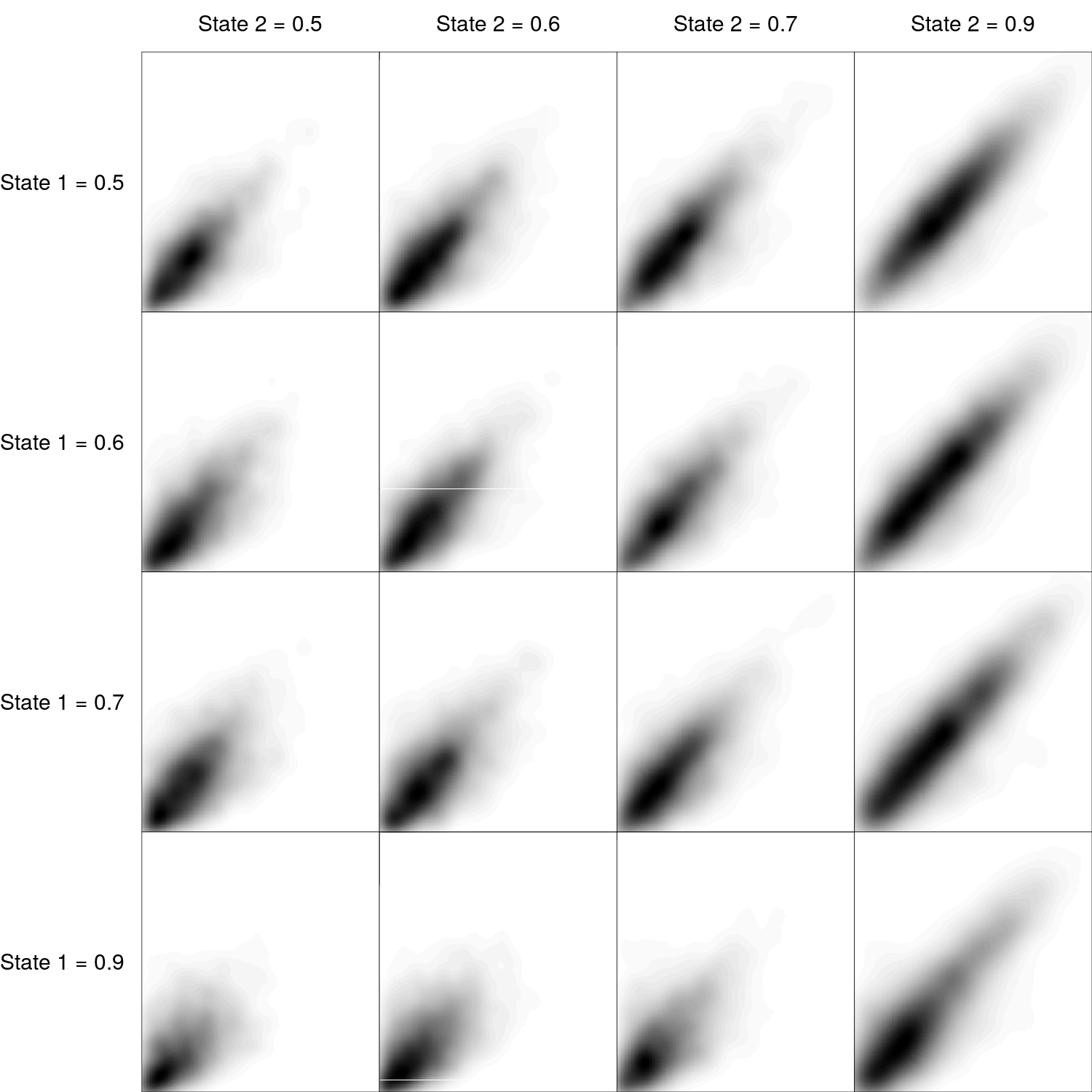}
   \caption{The row and column headings give the probability of staying within the given state. State 1 has low autocorrelation, state 2 has high autocorrelation. The density becomes more tightly spread along the line $y=x$ as the data remains in state 2 longer, as well as becoming less concentrated at any particular mean value.}
   \label{fig::simTransitlag}
\end{figure}

\begin{table}[H]
	\centering
	\begin{tabular}{rccrccrcc}
		\multicolumn{9}{c}{Two State Elk Parameters} \\
		$\mathbf{d_{(t,t+1)}}$		& State 1			& State 2			&	$\bm{\theta_{t}}$	& State 1			& State 2			& $\mathbf{A}$				& $\mathbf{p_{\cdot,1}}$& $\mathbf{p_{\cdot,2}}$\\
		\hline
		$\mathbf{\bm{\mu}_{RL,b}}$	& 3.364	& 0.355	& 	$\mathbf{c}$		& 0	& 0	& $\mathbf{p_{1,\cdot}}$	& 0.75		& 0.25	\\
		$\mathbf{\bm{\phi}_{b}}$	& 0	& 0	&	$\bm{\rho}$			& 0.228	& 0.6	& $\mathbf{p_{2,\cdot}}$	& 0.15		& 0.85	\\
		$\bm{\sigma}$				& 4.329	& 0.378	&						& 					&					& 							& 						& 				\\
	\end{tabular}
	\caption{Parameters for the two state elk simulations. Low autocorrelation is 0.1, medium autocorrelation is 0.4, high autocorrelation is 0.85. These parameters are slightly modified from the original source, the vignette to the R package moveHMM \cite{Michelot::moveHMM}.}
	\label{tab::twoElkPars}
\end{table}

\begin{table}[H]
	\centering
	\begin{tabular}{rcccrcccrccc}
		\multicolumn{12}{c}{Three State Seal Parameters} \\
		$\mathbf{d_{(t,t+1)}}$		& State 1			& State 2			& State 3			&	$\bm{\theta_{t}}$	& State 1			& State 2			& State 3			& $\mathbf{A}$				& $\mathbf{p_{\cdot,1}}$& $\mathbf{p_{\cdot,2}}$	& $\mathbf{p_{3,\cdot}}$ \\
		\hline
		$\mathbf{\bm{\mu}_{RL,b}}$	& 0.202	& 0.998	& 2.091	& 	$\mathbf{c}$		& 0	& 0	& 0	& $\mathbf{p_{1,\cdot}}$	& 0.848		& 0.142			& 0.005	\\
		$\mathbf{\bm{\phi}_{b}}$	& 0.04	& 0.429	& 0.945	&	$\bm{\rho}$			& 0.209	& 0.681	& 0.867	& $\mathbf{p_{2,\cdot}}$	& 0.065		& 0.754			& 0.164\\
		$\bm{\sigma}$				& 0.157	& 0.529	& 0.235	&						& 					&					& 					& $\mathbf{p_{3,\cdot}}$	& 0.005		& 0.164			& 0.831\\
	\end{tabular}
	\caption{Parameters for the three state seal simulations.}
	\label{tab::threeSealLengthPars}
\end{table}

\begin{table}[H]
	\centering
	\begin{tabular}{rccrccrcc}
		\multicolumn{9}{c}{Two State ``Best Practice" Seal Parameters} \\
		$\mathbf{d_{(t,t+1)}}$		& State 1			& State 2			&	$\bm{\theta_{t}}$	& State 1			& State 2			& $\mathbf{A}$				& $\mathbf{p_{\cdot,1}}$& $\mathbf{p_{\cdot,2}}$\\
		\hline
		$\mathbf{\bm{\mu}_{RL,b}}$	& 0.552	& 1.731	& 	$\mathbf{c}$		& 0	& 0	& $\mathbf{p_{1,\cdot}}$	& 0.826		& 0.174	\\
		$\mathbf{\bm{\phi}_{b}}$	& 0.407	& 0.892	&	$\bm{\rho}$			& 0.508	& 0.858	& $\mathbf{p_{2,\cdot}}$	& 0.12		& 0.88	\\
		$\bm{\sigma}$				& 0.351	& 0.244	&						& 					&					& 							& 						& 				\\
	\end{tabular}
	\caption{Parameters for the two state simulations. These parameters are estimated in the two state model of Section \ref{sec::mgreySeal}}
	\label{tab::twoSealPars}
\end{table}

\subsection{Graphics for the male grey seal}
\label{sapp::msealGraphics}

\noindent
\begin{figure}[H]
	\centering
	\includegraphics[width=\textwidth]{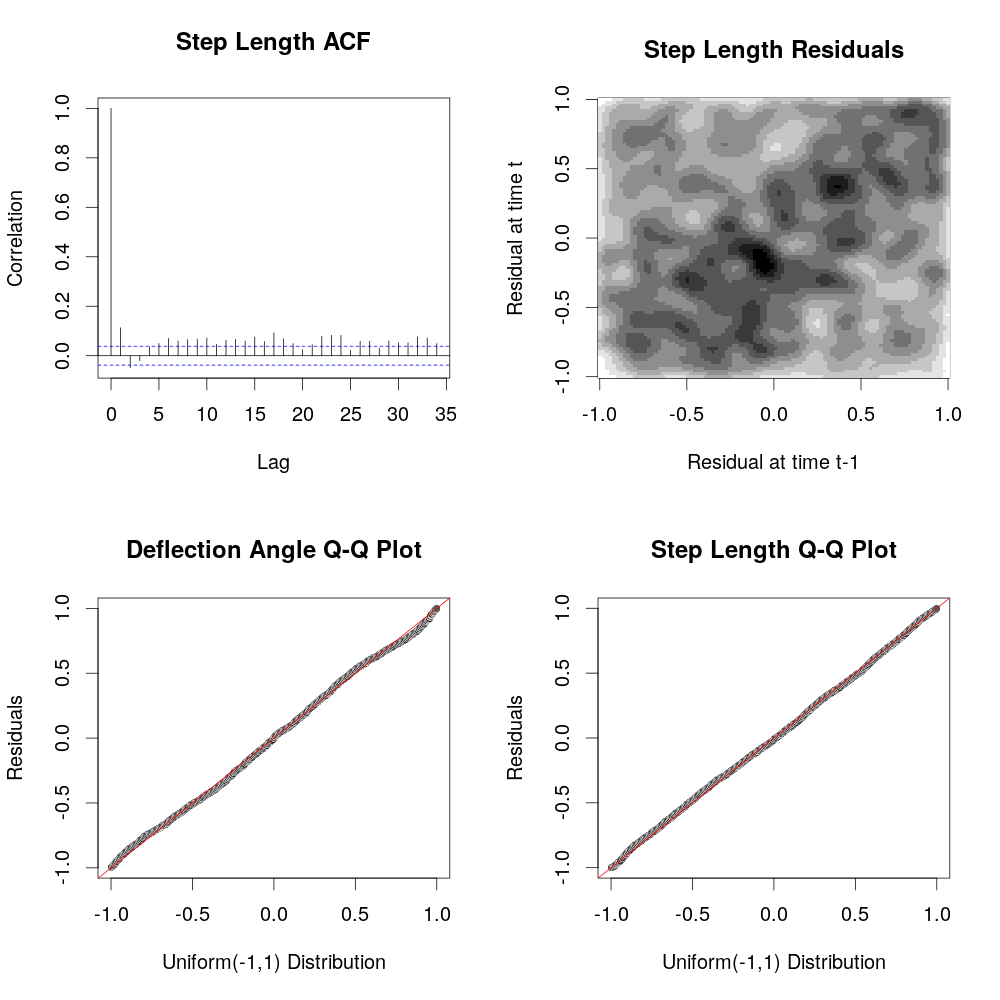}
	\caption{Step length lag-plot and residuals plots for the two state CarHMM.}
\end{figure}

\noindent
\begin{figure}[H]
	\centering
	\includegraphics[width=\textwidth]{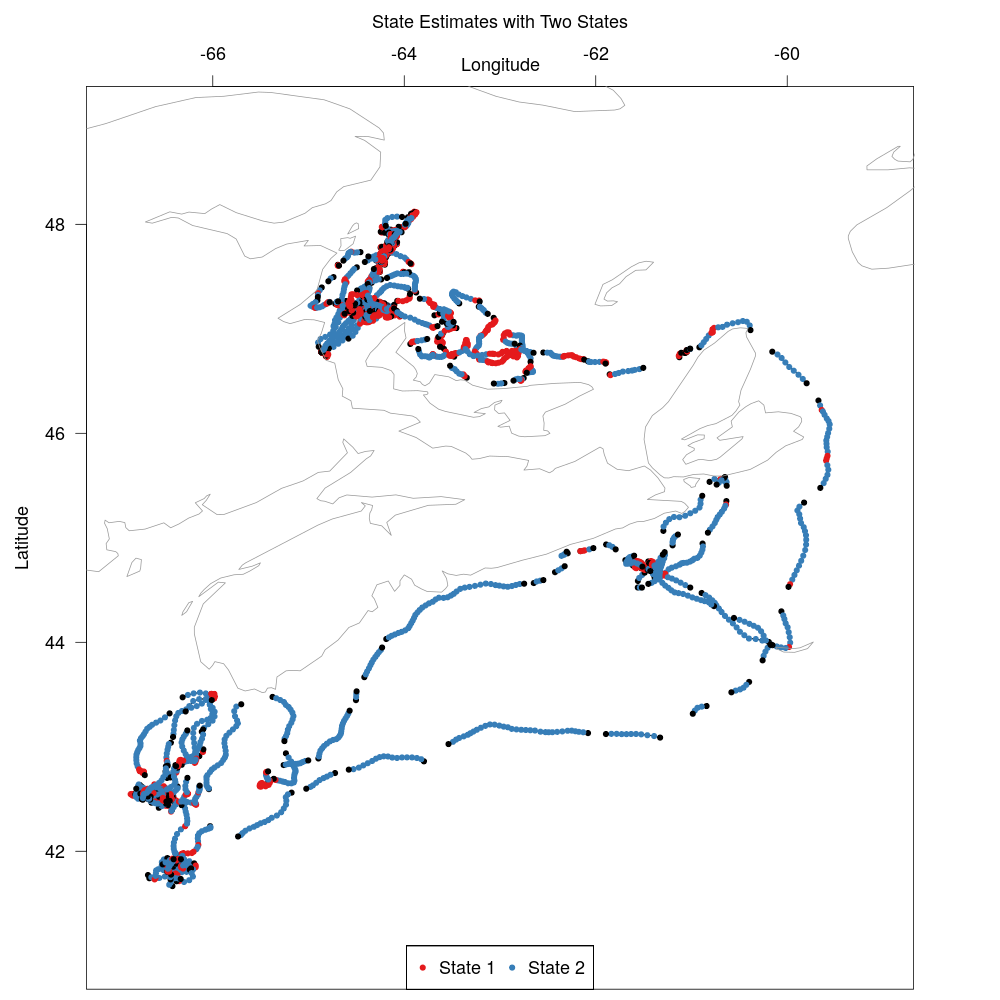}
	\caption{Map of the male grey seal track with state estimates from the two state CarHMM.}
\end{figure}

\noindent
\begin{figure}[H]
	\centering
	\includegraphics[width=\textwidth]{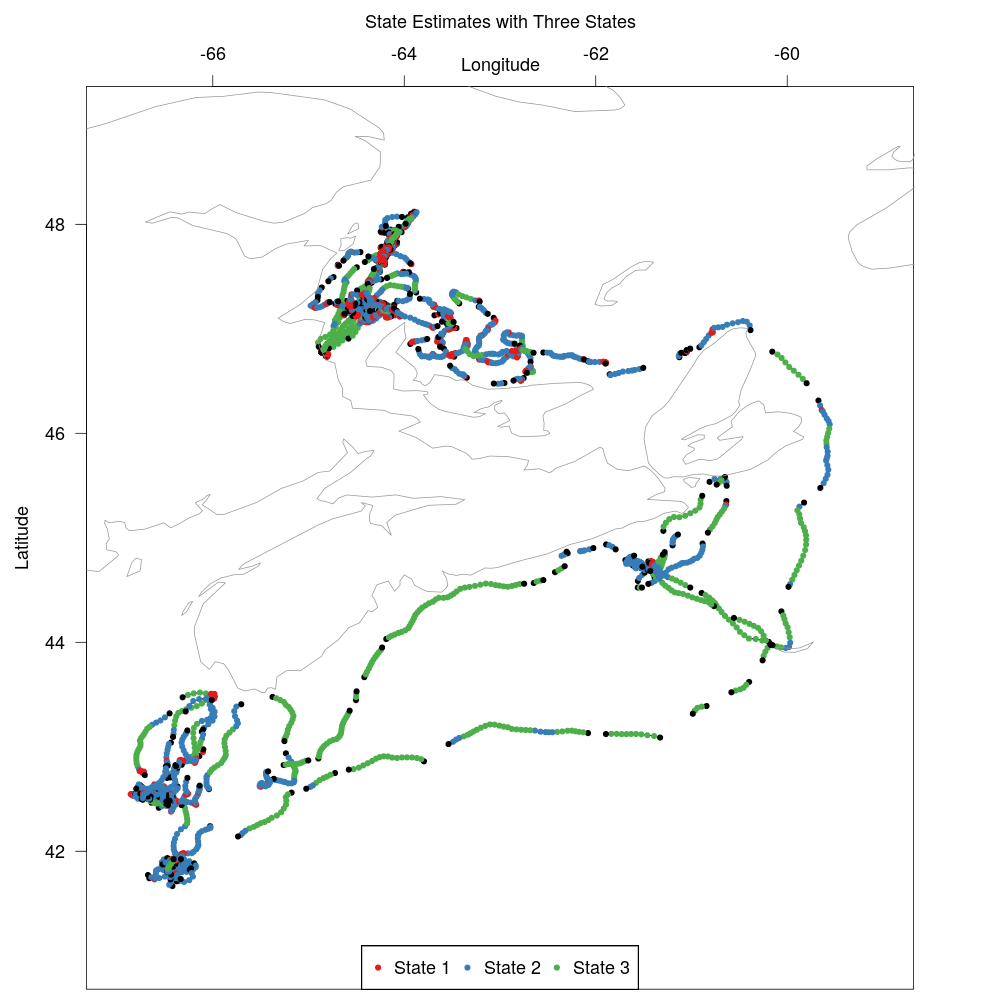}
	\caption{Map of the male grey seal track with state estimates from the three state CarHMM.}
\end{figure}

\end{document}